\documentclass[12pt,preprint]{aastex}

\shorttitle{Molecular Hydrogen in High-Velocity Clouds}
\shortauthors{Richter et al.}

\begin{document}

\title{Molecular Hydrogen in High-Velocity Clouds}


\author{Philipp Richter\altaffilmark{1}, 
Kenneth R. Sembach\altaffilmark{2,3}, 
Bart P. Wakker\altaffilmark{1}}

\and

\author{Blair D. Savage\altaffilmark{1}}


\altaffiltext{1}{Department of Astronomy, University of Wisconsin-Madison,
475 N. Charter Street, Madison, WI\,53706; richter@astro.wisc.edu}
\altaffiltext{2}{Department of Physics and Astronomy, Johns Hopkins University,
3400 N. Charles Street, Baltimore, MD\,21218}
\altaffiltext{3}{Current address: Space Telescope Science Institute, 3700 San Martin
Dr., Baltimore, MD\,21218}


\begin{abstract}

We present {\it Far Ultraviolet Spectroscopic Explorer} (FUSE)  
observations of interstellar molecular hydrogen (H$_2$) in 
two Galactic high-velocity clouds (HVCs). 
Molecular hydrogen absorption
is detected 
in the Magellanic Stream (abundance $\sim 0.3$ solar) 
toward the Seyfert galaxy Fairall\,9
in the lowest three rotational states ($J=0-2$)
at $v_{\rm LSR}=+190$ km\,s$^{-1}$,
yielding
a total H$_2$ column density of log $N$(H$_2$)$=16.40^{+0.26}_{-0.53}$.
In contrast, no H$_2$ absorption is seen in the high-velocity
cloud Complex C (abundance $\sim 0.1$ solar)
toward the
quasar PG\,1259+593 (log $N$(H$_2$)$\le 13.96$ at $v_{\rm LSR}=-130$ km\,s$^{-1}$), 
although both HVCs
have similar H\,{\sc i} column densities on the order
of log $N$(H\,{\sc i})$\approx 20$. 
Weak H$_2$ absorption is detected in the Intermediate-Velocity
Arch (IV Arch; abundance $\sim 1.0$ solar) toward PG\,1259+593 (log $N$(H$_2$)$=14.10^{+0.21}_{-0.44}$
at $v_{\rm LSR}=-55$ km\,s$^{-1}$ and log $N$(H\,{\sc i})
$=19.5$).
It thus appears that metal- and dust-poor halo clouds like
Complex C are not able to form and maintain widely distributed
H$_2$, whereas metal and dust-rich halo clouds like the IV Arch can maintain
H$_2$ even at low H\,{\sc i} column densities.
\end{abstract}


\keywords{ISM: clouds -- ISM: abundances -- quasars: absorption lines -- 
quasars: individual (Fairall\,9, PG\,1259\,+593) -- Galaxy: halo}


\section{Introduction}

The {\it Far Ultraviolet Spectroscopic Explorer} (FUSE)
is the first instrument to be used to systematically
study atomic and molecular absorption lines in Galactic halo clouds
in the important far-ultraviolet (FUV) spectral range ($\lambda < 1150$ \AA).
These halo clouds are seen in H\,{\sc i} 21cm emission 
at radial velocities that do not match a simple model of differential 
Galactic rotation. A separation is traditionally made between high-velocity clouds (HVCs;
$|V_{\rm LSR}|>90$ km\,s$^{-1}$) and intermediate-velocity clouds (IVCs;
$30$ km\,s$^{-1} < |V_{\rm LSR}| < 90$ km\,s$^{-1}$). 
The presence of molecular material would give important new
insights into the physical conditions in the interiors of these clouds, and
would allow the study of molecular gas under conditions that are likely to
be very different from those in the disk of the Milky Way. In addition,
it has been proposed that dense molecular regions in the Galactic halo
could serve as birth places for the population of young B-type stars found in
the Milky Way halo (Conlon et al.\,1992), or even as candidates for baryonic dark
matter (e.g., de\,Paolis et al.\,1995; Kalberla et al.\,2000).

Until recently, most searches for molecular gas in HVCs were restricted to
observations of CO (e.g., Wakker et al.\,1997; Akeson \& Blitz 1999),
but neither CO emission nor absorption was found.
In their survey of HCO$^+$ absorption in HVCs
toward 27 quasars, Combes \& Charmandaris (2000) reported one tentative
detection, but this case has not been confirmed yet.
The most abundant molecule in the Universe, molecular hydrogen (H$_2$), could
not be observed in HVCs and IVCs before 1996 because of the lack of suitable 
space-based instrumentation to
study H$_2$ absorption in the Galactic halo
at wavelengths between 900 and 1130 \AA. With new
FUV instruments such as FUSE and the {\it Orbiting and Retrievable Far and Extreme
Ultraviolet Spectrometer} (ORFEUS), this wavelength
range has now become accessible, and recent studies of various 
Galactic halo cloud complexes have unveiled the presence of diffuse H$_2$ 
in both HVCs (Richter et al.\,1999; Sembach et al.\,2001) and
IVCs (e.g., Gringel et al.\,2000; Richter et al.\,2001a), albeit
at low column densities (log $N$(H$_2$)$\le 17$). While diffuse H$_2$
appears to be a rather widespread constituent in the more nearby
and metal-rich IVCs, H$_2$ in HVCs has been detected
only in two cases (Richter et al.\,1999; Sembach et al.\,2001).  
However, H$_2$ measurements of gas in HVCs are more difficult 
due to the limited availability of suitable background sources
at large distances from the Galactic plane.

\section{The Magellanic Stream and Complex C}

In this paper we investigate the molecular hydrogen content in the Magellanic Stream
toward the Seyfert galaxy Fairall\,9 ($l=295\fdg1, b=-57\fdg8$; $V=13.83$; $z=0.047$)
and in high-velocity cloud Complex C toward the quasar PG\,1259+593 
($l=120\fdg6, b=+58\fdg1$; $V=15.84$; $z=0.478$). 
Figure 1 shows the location of these two sight lines plotted on the Magellanic Stream
and Complex C H\,{\sc i} 21cm
emission maps (Hulsbosch \& Wakker 1988; Bajaja et al.\,1985; Morras et al.\,2000).
The Magellanic Stream has a metallicity of $\sim 0.3$ solar (Gibson et al.\,2000;
Lu et al.\,1998) and is 
believed to be tidally torn out of the Small Magellanic Cloud (SMC). 
Sembach et al.\,(2001) have detected H$_2$ absorption in the Leading Arm of the Magellanic
Stream toward the Seyfert galaxy NGC\,3783 with log $N$(H$_2$)$=16.80 \pm 0.10$.
Toward Fairall\,9, Parkes 21cm emission line data ($17 \farcm 0$ beam) show two 
blended H\,{\sc i} components 
at $v_{\rm LSR}=+149$ and $+195$ km\,s$^{-1}$ with a total integrated H\,{\sc i} column density
of log $N=19.97 \pm 0.01$ (Gibson et al.\,2000).
For Complex C,
Wakker et al.\,(1999) and Richter et al.\,(2001b) find elemental abundances of $\sim 0.1$ solar,
suggesting that Complex C represents the infall of intergalactic material onto the Milky Way,
but Gibson et al.\,(2001) reported abundances for Complex in other directions that vary between
0.08 and 0.44 solar.
Murphy et al.\,(2000) did not find H$_2$ absorption in Complex C toward Mrk 876 (log $N$(H$_2$)$\le 14.30$).
Toward PG\,1259+593, Effelsberg 21cm data ($9\farcm1$ beam) show Complex C H\,{\sc i} 
21cm emission at $v_{\rm LSR}=-130$ km\,s$^{-1}$ with log $N$(H\,{\sc i})$=19.92 \pm 0.01$
(Wakker et al.\,2001), but other instruments at lower resolution (see Richter et al.\,2001b)
yield lower column densities observing the same direction, indicating that there is sub-structure
in Complex C on $10-20$ arc minute scales.
The H\,{\sc i} profiles toward PG\,1259+593 (sampling Complex C) and Fairall\,9 (sampling
the Magellanic Stream)
are shown in the top panels of Figure 2.
 
\section{FUSE Observations and Data Analysis}

Four optical channels are available on FUSE, two SiC channels from 905 to 1100 \AA, and
2 LiF channels covering 1000 to 1187 \AA\, (for instrument descriptions and performance
information see Moos et al.\,(2000)
and Sahnow et al.\,(2000)). 
FUSE observations of Fairall\,9 were conducted 7 July 2000, and observations of PG\,1259+593 
were performed between February 2000 and
March 2001. The observations were obtained using the large aperture (LWRS) and
the photon address mode,
providing spectra at a resolution of $\sim 25$ km\,s$^{-1}$ (FWHM).
Total integration times were $\sim 35$ ks (Fairall\,9), and $\sim 400$ ks
\footnote{After correction for event bursts in the raw data.}
(PG\,1259+593).
The data were reduced with the CALFUSE (v.1.8.7) calibration pipeline and 
rebinned to $8$ km\,s$^{-1}$ wide pixels. 
The wavelength 
calibration (accurate to approximately $\pm 10$ km\,s$^{-1}$) is based on 
aligning various atomic absorption
lines with the H\,{\sc i} 21cm emission data.

For PG\,1259+593, the average flux in the FUSE spectrum
is $\sim 2 \times 10^{-14}$ erg\,cm$^{-2}$\,s$^{-1}$\,\AA$^{-1}$. 
The flux is almost constant over the wavelength range sampled by FUSE, and 
the typical signal-to-noise
ratio (S/N) is $\sim 17$ per pixel element after rebinning. 
The average flux in the Fairall\,9 spectrum is 
also $\sim 2 \times 10^{-14}$ erg\,cm$^{-2}$\,s$^{-1}$\,\AA$^{-1}$,
so that with $t_{\rm obs}=35$ ks the S/N is $\sim 3$ 
at most wavelengths. This is too low
to accurately measure absorption lines with equivalent widths ($W_{\lambda }$) less than
$\sim 80$ m\AA\, at wavelengths $< 1040$ \AA, where numerous atomic
and molecular lines cause severe blending problems. 
For $\lambda > 1040$ \AA, however, the flux in the Fairall\,9 spectrum
rises to $\sim 8 \times 10^{-14}$ erg\,cm$^{-2}$\,s$^{-1}$\,\AA$^{-1}$ due
to intrinsic Ly\,$\beta$ emission from Fairall\,9 itself. This allows us to study molecular hydrogen
toward Fairall\,9 in the well-separated Lyman $0-0$, $2-0$, $3-0$ and $4-0$ bands 
at an average S/N of $\sim 6$ per rebinned pixel element.  
For the following analysis we used data from the SiC\,2A, LiF\,1A, and LiF\,1B data segments,
which provide higher S/N and better resolution than the SiC\,1B, LiF\,2B, and LiF\,2A
data.

\section{H$_2$ Measurements}

In the spectrum of Fairall\,9, we find H$_2$ absorption in lines from the rotational
levels $J=0,1$, and $2$ 
in the Magellanic Stream at $v_{\rm LSR}=+190$ km\,s$^{-1}$.
We have selected 14 lines that have sufficient
S/N for a reliable analysis.
Low-order polynomials were fit to the continua and
equivalent widths were measured by fitting single component Gaussians to the H$_2$ lines.
H$_2$ absorption line profiles are shown in Figure 2, and equivalent widths are listed
in Figure 3. There is an indication that the H$_2$ line profiles have a second weak
component at $v_{\rm LSR}=+149$ km\,s$^{-1}$, similar to the H\,{\sc i} emission pattern.
The data quality, however, is not good enough to separate these two components,
and the observed equivalent widths are clearly dominated by the $+190$ km\,s$^{-1}$ absorption.
Almost no H$_2$ absorption is seen in the local Galactic gas at $0$ km\,s$^{-1}$.
H$_2$ column densities for the Magellanic Stream were derived by 
fitting the absorption lines from the individual rotational states
to a curve of growth with $b=5.0^{+3.4}_{-1.2}$ km\,s$^{-1}$, which represents the best 
fit to the data. We obtain logarithmic column densities, log $N(J)$, of
log $N(0)=15.86^{+0.40}_{-0.46}$, log $N(1)=16.22^{+0.21}_{-0.54}$, log $N(2)=15.04^{+0.28}_{-0.23}$,
and log $N(3) \le 14.45$ ($3\sigma$).
The total H$_2$ column density is log $N=16.40^{+0.28}_{-0.53}$, and the average fraction of 
hydrogen in molecular form is $f=2N$(H$_2$)/[$N$(H\,{\sc i})$+2N$(H$_2$)]$=5.4 \times 10^{-4}$,
using the total H\,{\sc i} column density for the Magellanic Stream from the 
Parkes data shown in Figure 2. We derive an excitation temperature
of T$_{\rm ex}=142 \pm 30$ K by fitting the rotational level populations of $J=0, 1$, and $2$ 
to a single Boltzmann distribution (see Figure 3).
\footnote{The single temperature fit indicates that the ortho-to-para H$_2$ ratio
is in local thermodynamical equilibrium.}
The temperature 
is very similar to that found in the Leading Arm of
the Magellanic Stream
by Sembach et al.\,(2001; T$_{\rm ex}=133^{+37}_{-21}$ K). Possibly,
both values represent the kinetic temperature of the
gas in which the H$_2$ resides. These
temperatures are roughly twice as
high as the kinetic temperatures found in the
Milky Way disk (Savage et al.\,1977), but are more similar to
those found for high-latitude clouds (Shull et al.\,2000).   

Toward PG\,1259+593, atomic absorption is present at Complex C 
velocities near $v_{\rm LSR}=-130$ km\,s$^{-1}$ (Richter et al.\,2001b),
but no significant H$_2$ absorption is detected (Figure 2).
\footnote{The only noteworthy H$_2$ feature seen at Complex C velocities is that in the
Werner Q(2),0-0 line (see Fig.\,2), but this absorption is probably a contaminating
intergalactic absorber or an instrumental
artifact since there is no
H$_2$ absorption seen in other $J=2$ transitions.}
We analyze the strongest of the H$_2$ absorption
lines in the Werner band $J=0$ and $1$ rotational levels 
at $\lambda = 985.6$ and $1008.6$ \AA, deriving $3\sigma$
detection limits on the order of 25 m\AA. 
From that we determine a $3 \sigma$ upper limit of log $N$(H$_2$)$\le 13.96$ 
for the total H$_2$ column density in Complex C, assuming that these
lines lie on the linear part of the curve of growth. 
Using the Effelsberg H\,{\sc i} data,
we find $f$(H$_2$)$\le 2.2 \times 10^{-6}$ ($3 \sigma$).
While the FUSE data of PG\,1259+593 show no evidence for H$_2$ 
in Complex C, very weak H$_2$ absorption is detected in 6
$J=0-2$ lines at $v_{\rm LSR}=-55$ km\,s$^{-1}$ (see Figure 2), 
related to gas of the Intermediate Velocity Arch (IV Arch) in the lower
Galactic halo (see Richter et al.\,2001b). This finding is consistent with a previous detection of
H$_2$ in this IVC toward the halo star HD\,93521 (Gringel et al.\,2000)
in roughly the same direction of the sky. 
For this component, we find a total H$_2$ column density of
log $N=14.10^{+0.21}_{-0.44}$, assuming that the lines fall on the
linear part of the curve of growth.
\footnote{Because of the small number of lines and the resulting large uncertainties 
for the individual column densities, $N(J)$, a reliable estimate for 
T$_{\rm ex}$ is not possible.}

\section{Discussion}

Molecular hydrogen absorption in 
HVCs has now been detected
in three HVC sight lines.
The first detection of H$_2$ in HVC gas was reported by
Richter et al.\,(1999) for the high-velocity gas in front
of the Large Magellanic Cloud (LMC) toward HD\,269546. The second detection was
that of Sembach et al.\,(2001), who found H$_2$ in the
Leading Arm of the Magellanic Stream in the direction of NGC\,3783.
The results presented here show that H$_2$ is also present
in the main body of the Magellanic Stream, but probably
not in Complex C.

A preliminary analysis of more than 100 FUSE spectra of
quasars, AGNs, and halo stars (including several sightlines
passing through Complex C) gives also no evidence
for the existence of H$_2$ in HVCs other than 
the Magellanic Stream and in
the cloud in front of the LMC, but shows the presence
of H$_2$ absorption in IVCs in at least 15 spectra 
(H$_2$ detections include the IV Arch, the LLIV Arch, Complex gp,
and the IV Spur; see Richter 2001c). 
It appears that diffuse H$_2$ is rather widespread in
IVCs, but present only in certain HVCs.
Most likely, this 
is a metallicity effect:
IVCs tend to have nearly solar abundances (e.g., Richter et al.\,2001a),
while abundances in HVCs
are as low as $\sim 0.1$ solar (e.g. Complex C; 
Wakker et al.\,1999; Richter et al.\,2001b).
The abundances and distances measured in IVCs so far 
support the idea that they represent the
returning, cooled gas of a ``Galactic fountain'' 
(Houck \& Bregman 1990). If so, the H$_2$ found in IVCs must 
have formed {\it in} the halo, because it is 
unlikely that molecular material survives
the violent processes that 
ejects the gas into the halo.
A fraction of the available heavy elements
is incorporated into
dust grains, on whose surface the H$_2$ formation
proceeds most efficiently (e.g., Pirronello et al.\,1999).
Thus, metal and dust-deficient clouds should have much lower H$_2$ formation rates
than those with higher abundances.
The depletion of Fe\,{\sc ii} and Si\,{\sc ii} in the IV Arch
(Richter et al.\,2001b), in the Magellanic Stream (see Sembach et al.\,2001),
and in the HVC in front of the LMC
indicates that these clouds contain dust grains, in contrast
to Complex C, where the abundance pattern suggests that
Complex C contains little, or no dust at all
(Richter et al.\,2001b).
The detections of H$_2$
in IVCs and HVCs that contain heavy elements 
in comparison to the
H$_2$ non-detections in metal- and dust-deficient clouds like 
Complex C therefore suggest that the H$_2$ formation in
Galactic halo clouds is very sensitive to the
dust abundance.
Metal- and dust-deficient clouds like Complex C are
probably not able to form and maintain widely
distributed H$_2$ gas.
However, if the molecular material in metal-poor halo clouds 
is highly concentrated in small, dense clumps, where the
H$_2$ formation rate (increasing with density)
can compete
with the photo-dissociation (which is
reduced by H$_2$ line
self-shielding), it may remain hidden from observation.

In the case of the Magellanic Stream, Sembach et al. (2001)
have suggested that the H$_2$ may have formed in the
Magellanic system and has survived the tidal stripping.
In view of the overall presence of H$_2$ in IVCs, however,
it is just as likely that the H$_2$ has formed in situ
on dust grains during the 2 Gyr orbital period
of the Magellanic Stream.
Possibly, H$_2$ in the Magellanic Stream and in
IVCs forms quickly in small, 
compact cloudlets
and then is dispersed into a larger volume by cloudlet
collisions or other disruptive processes. The diffuse
molecular gas in the Magellanic Stream and in
IVCs then may trace a more
dense (but yet undetected) molecular gas phase in the halo 
in which star formation might occur
(see Dyson \& Hartquist 1983; Conlon et al.\,1992).

Clearly, more FUV absorption line measurements are desirable
to further investigate the molecular gas phase in
the Galactic halo.
A systematic study of H$_2$ in IVCs
and HVCs 
could also help to characterize the formation
and dissociation processes for H$_2$ in diffuse gas in
environments at low metallicities and moderate FUV
radiation fields, such as in low surface brightness galaxies
and intergalactic H\,{\sc i} clouds.

\acknowledgments

This work is based on data obtained for the
the Guaranteed Time Team by the NASA-CNES-CSA FUSE
mission operated by the Johns Hopkins University.
Financial support has been provided by NASA
contract NAS5-32985.

\newpage

\resizebox{1.00\hsize}{!}{\includegraphics{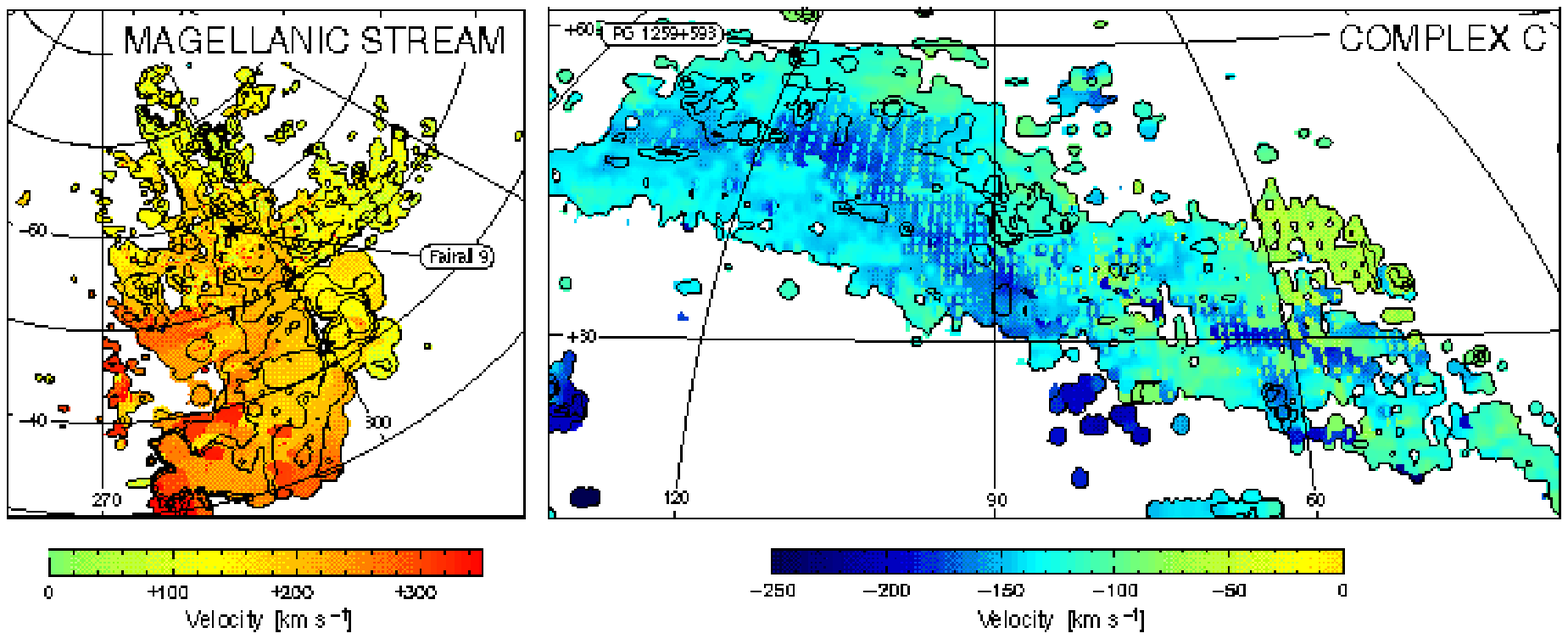}}
\figcaption[f1.eps]{
H\,{\sc i} 21cm maps ($l$, $b$)
of the Magellanic Stream (MS; left panel) and
Complex C (right panel; for references see text).
The sight lines toward Fairall\,9 (MS) and PG\,1259+593 (Complex C) are
labeled in the plot. Contours represent H\,{\sc i} column densities of
$0.3, 2.0, 6.5$ and $13.0 \times 10^{19}$ cm$^{2}$. The grey (colour) scale shows velocity
[A higher quality version of this Figure is available on request].
\label{fig1}}

\resizebox{0.65\hsize}{!}{\includegraphics{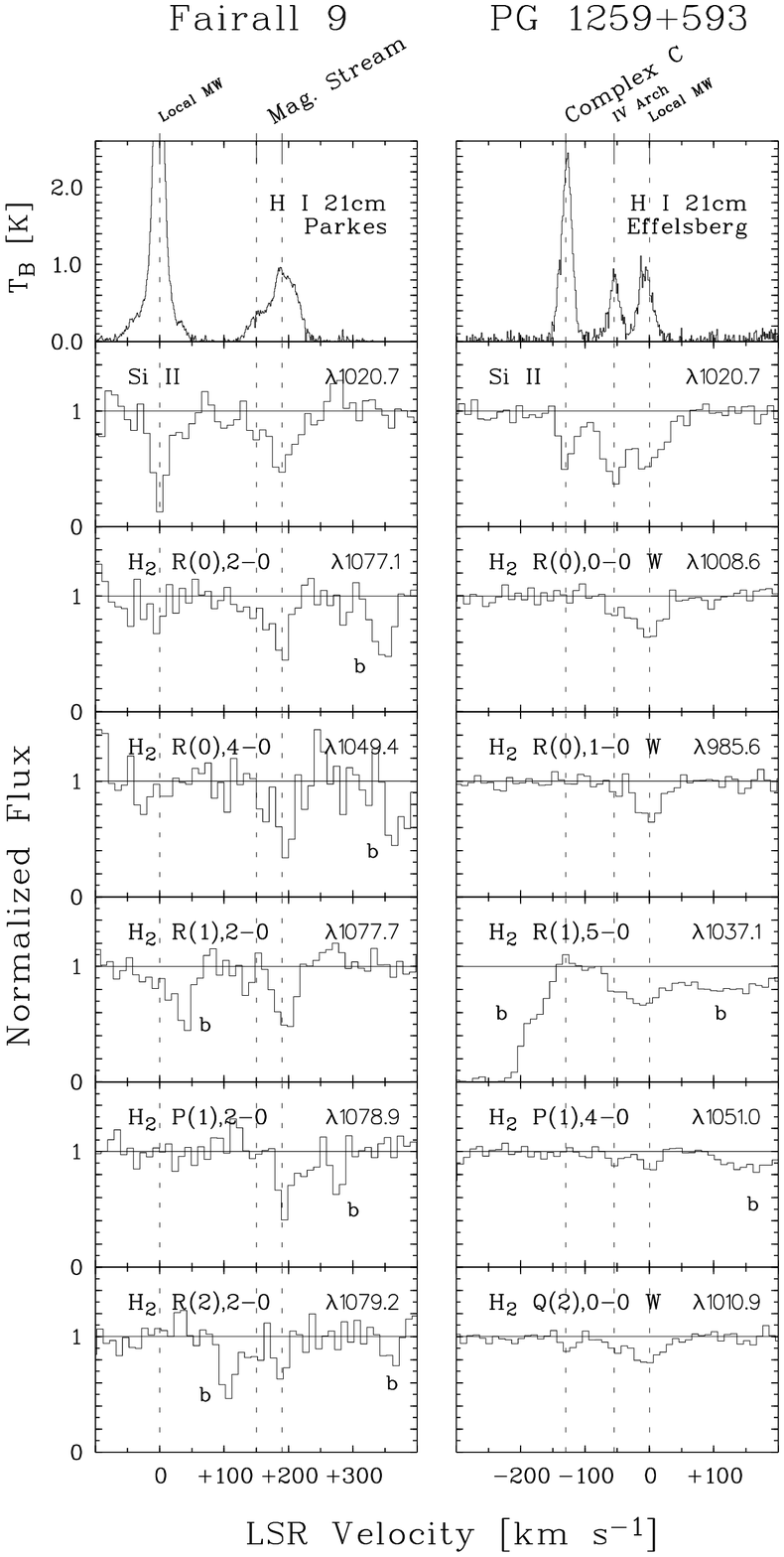}}
\newpage
\figcaption[f2.eps]{
Interstellar H$_2$ absorption line profiles from the FUSE spectra of Fairall\,9
(left panel) and PG\,1259+593 (right panel).
H\,{\sc i} emission line spectra from Parkes and
Effelsberg are plotted in the uppermost box. For
comparison, Si\,{\sc ii} $\lambda 1020.7$ line profiles are also shown. The
various absorption components are labeled above the boxes;
H$_2$ transitions from the Werner band are
labeled with `W'.
In the spectrum of Fairall\,9, H$_2$ absorption
in the Magellanic Stream is clearly visible at
$v_{\rm LSR}=+190$ km\,s$^{-1}$. No H$_2$ is seen in Complex C at
$v_{\rm LSR}=-130$ km\,s$^{-1}$ in the spectrum of PG\,1259+593,
but weak H$_2$ absorption is present in the IV Arch component
at $v_{\rm LSR}=-55$ km\,s$^{-1}$. Blending lines from other species are
marked with `b'.
\label{fig2}}

\newpage
\resizebox{0.70\hsize}{!}{\includegraphics{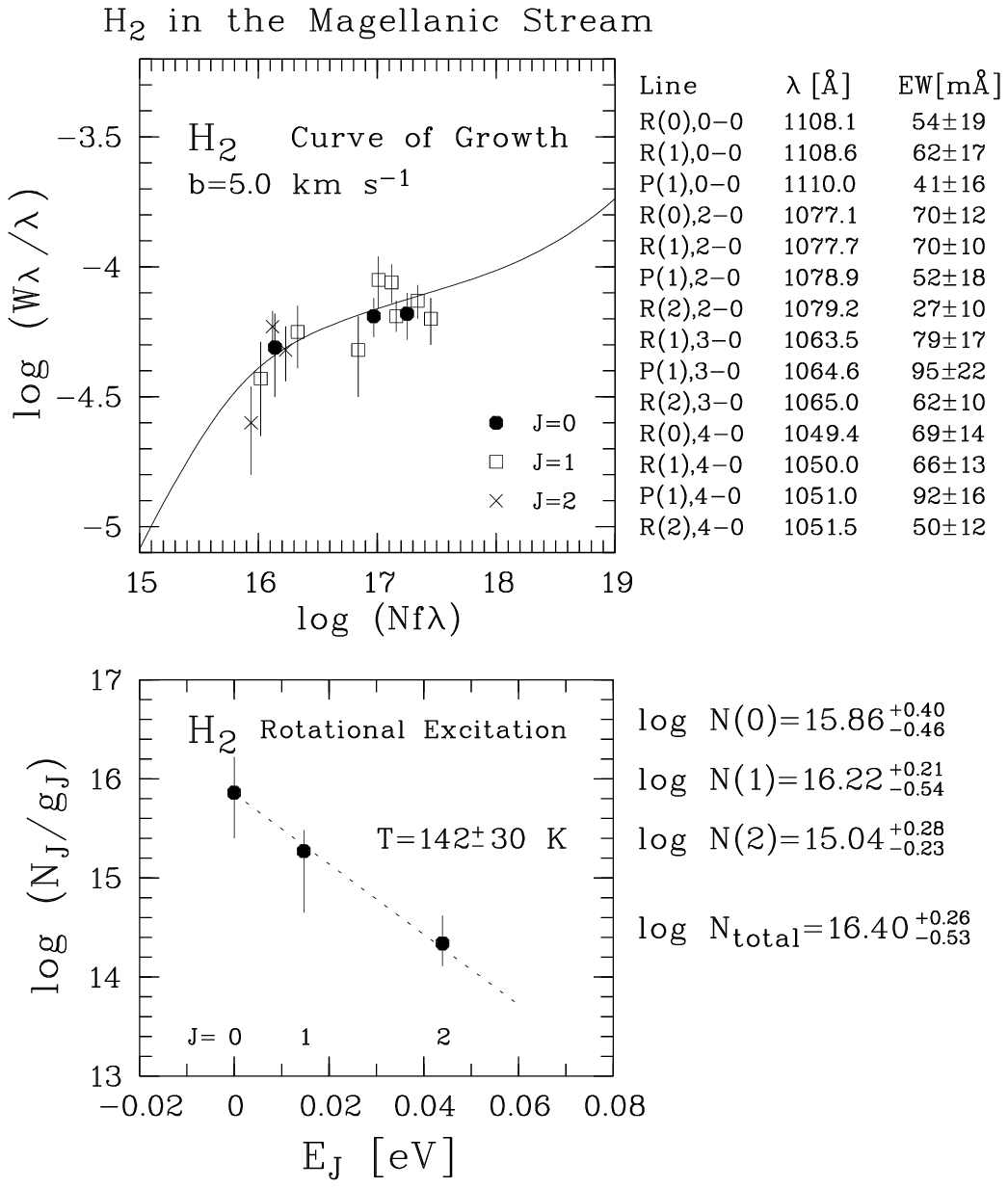}}
\figcaption[f3.eps]{
Empirical curve of growth for the measured H$_2$ absorption
lines in the Magellanic Stream toward Fairall\,9 (upper panel).
Wavelengths and equivalent widths of these lines are listed
on the right-hand side next to the plot. The lower panel
shows the rotational excitation of the H$_2$ gas in the
Magellanic Stream, equivalent to a Boltzmann temperature
of T$_{\rm ex}=142 \pm 30$ K. H$_2$ column densities
are given on the right-hand side next to the plot.
\label{fig3}}

\end{document}